\documentclass[cameraready]{Interspeech}
\title{MUGEN: Evaluating and Improving Multi-audio Understanding of Large Audio-Language Models}

\author[affiliation={1}, equalcontribution]{Chih-Kai}{Yang}
\author[affiliation={1}, equalcontribution]{Yun-Shao}{Tsai}
\author[affiliation={2, \dagger}]{Yu-Kai}{Guo}
\author[affiliation={2, \dagger}]{Ping-Le}{Tsai}
\author[affiliation={2, \ddagger}]{Yen-Ting}{Piao}
\author[affiliation={2, \ddagger}]{Hung-Wei}{Chen}
\author[affiliation={2, \ddagger}]{Ting-Lin}{Hsiao}
\author[affiliation={2, \ddagger}]{Yun-Man}{Hsu}
\author[affiliation={1}]{Ke-Han}{Lu}
\author[affiliation={3}, correspondingauthor]{Hung-yi}{Lee}

\address{
  $^1$ Graduate Institute of Communication Engineering, National Taiwan University, Taiwan \\
  $^2$ National Taiwan University, Taiwan \\
  $^3$ NTU Artificial Intelligence Center of Research Excellence (NTU AI-CoRE), Taiwan
}

\email{chihkaiyang1124@gmail.com, r14942093@ntu.edu.tw, hungyilee@ntu.edu.tw}

\keywords{large audio-language model, multi-audio understanding, benchmark}

\usepackage{comment}
\usepackage{cite}
\usepackage{hyperref}
\usepackage{url}
\usepackage{makecell}
\usepackage{multirow}
\usepackage{xcolor}
\usepackage{relsize}
\usepackage{graphicx}
\usepackage{subcaption}
\usepackage{caption}

\begin{document}

\maketitle

\begin{abstract}
% While multi-audio reasoning is critical for large audio-language models (LALMs), it remains underexplored. We introduce MUGEN, a comprehensive benchmark evaluating this capability across speech, general audio, and music. Our experiments reveal consistent weaknesses in multi-audio settings, and performance degrades sharply as the number of concurrent audio inputs increases, identifying candidate scaling as a fundamental bottleneck. To mitigate this, we explore Audio-Permutational Self-Consistency (APSC), a training-free strategy that diversifies the presentation order of audio candidates. This approach helps models form a more robust aggregated prediction, achieving up to 6.74\% accuracy gains. Furthermore, combining this permutation strategy with Chain-of-Thought reasoning yields further performance improvements. These results expose fundamental blind spots in current LALMs and establish a critical foundation for evaluating complex auditory comprehension.

While multi-audio understanding is critical for large audio-language models (LALMs), it remains underexplored. We introduce MUGEN, a comprehensive benchmark evaluating this capability across speech, general audio, and music. Our experiments reveal consistent weaknesses in multi-audio settings, and performance degrades sharply as the number of concurrent audio inputs increases, identifying input scaling as a fundamental bottleneck. We further investigate training-free strategies and observe that Audio-Permutational Self-Consistency, which diversifies the order of audio candidates, helps models form more robust aggregated predictions, yielding up to 6.28\% accuracy gains. Combining this permutation strategy with Chain-of-Thought further improves performance to 6.74\%. These results expose blind spots in current LALMs and provide a foundation for evaluating complex auditory comprehension.
\end{abstract}

\section{Introduction}

Large language models (LLMs)~\cite{hurst2024gpt, grattafiori2024llama, yang2025qwen3} have achieved remarkable progress in language understanding and have been extended to multimodal domains such as vision~\cite{vipergpt} and speech~\cite{huang2023audiogpt, copilot}.
Building on this trend, large audio-language models (LALMs)~\cite{desta25, desta2, tp1, lin2025preliminary, qwen25omni, af3, voxtral, phi4} integrate auditory perception~\cite{yang2025audiolens, sake} with strong reasoning capability, enabling flexible interfaces for audio-centric applications such as voice agents~\cite{lyra}.
However, current research predominantly evaluates these models in isolated and single-audio environments.

In real-world deployments, LALMs require reasoning over \emph{multiple} audio segments simultaneously. This requirement appears in settings such as audio-based in-context learning~\cite{speech-icl}, where multiple audio demonstrations are required to guide adaptation, as well as in applications including speech retrieval-augmented generation (RAG)~\cite{speech-rag}, multi-speaker analytics, and cross-utterance event matching. These scenarios all involve jointly understanding multiple audio segments, requiring models to compare, aggregate, and reconcile information across clips.
Consequently, multi-audio understanding is not merely an advanced feature but a strict prerequisite for practical LALMs.

Despite its importance, current evaluation remains largely focused on single-audio settings. Existing benchmarks~\cite{yang-etal-2025-towards-holistic} assess general understanding~\cite{airbench, audiobench}, reasoning~\cite{sakura, mmaupro, speechifeval}, dialogue~\cite{sdeval}, bias~\cite{listenspeakfairly}, and safety~\cite{achilles, feng2025investigating}, but they largely overlook multi-audio scenarios. Recent attempts at multi-audio evaluation still exhibit two limitations: (1) narrow coverage of auditory attributes~\cite{chen-etal-2024-beyond-single, polyaudio, adiff}, often emphasizing semantic content or sound events while underrepresenting non-semantic aspects such as emotion; and (2) limited input scale~\cite{dynamicsuperb, mmaupro, adiff, mellow, cszs}, typically involving only two to three audio clips per sample. As a result, systematic evaluation across diverse auditory dimensions and larger input scales remains underexplored.

\begin{figure*}[t]
  \centering
  \includegraphics[width=\linewidth]{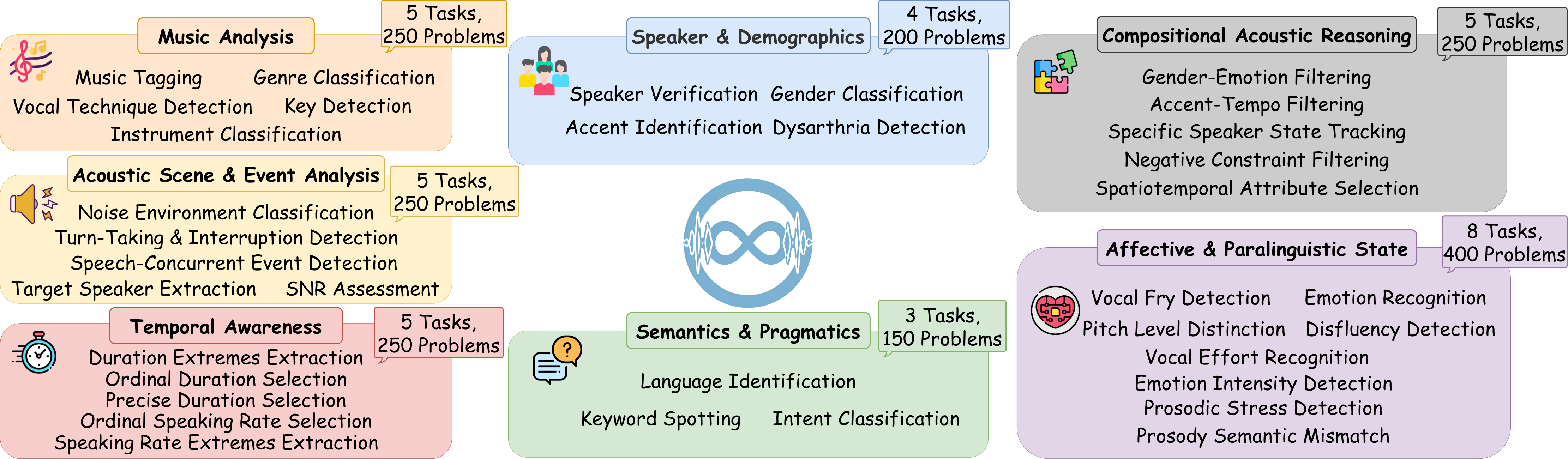}
  \caption{Overview of MUGEN and the detailed task distribution across the seven evaluation dimensions.}
  \label{fig:lag}
\end{figure*}

\begin{figure}[t]
  \centering
  \includegraphics[width=0.7\linewidth]{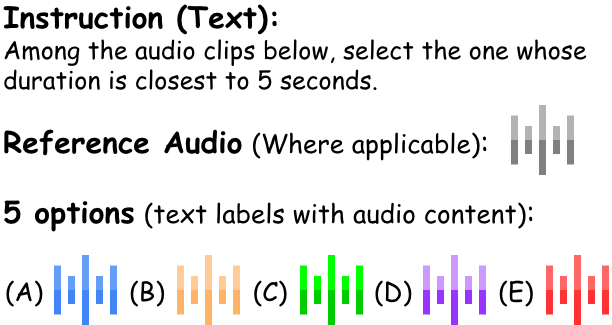}
  \vspace{-2mm}
  \caption{Illustration of the audio-as-option design.}
  \label{fig:format}
\end{figure}

To fill this gap, we introduce \textbf{MUGEN} (\underline{{Mu}}lti-audio \underline{{G}}rounding and Und\underline{{e}}rsta\underline{{n}}ding Benchmark), comprising 35 audio-grounding tasks across seven dimensions spanning speech, general audio, and music (Figure~\ref{fig:lag}). Each task requires selecting the audio that best satisfies a constraint from five candidates, enforcing cross-audio comparison through an audio-as-option design where all choices are audio signals rather than text (Figure~\ref{fig:format}). This design requires direct comparison of auditory features across candidates, introducing different challenges from text-based multiple-choice settings. Compared to prior work, MUGEN scales to more audio inputs and covers diverse auditory dimensions, enabling fine-grained multi-audio evaluation. Crucially, by emphasizing non-semantic and paralinguistic dimensions, MUGEN prevents models from bypassing acoustic reasoning through semantic shortcuts.

By benchmarking seven advanced LALMs, we uncover consistent weaknesses in multi-audio settings, especially for non-semantic attributes. Through systematic evaluation across varying numbers of audio inputs, we demonstrate that performance degrades markedly as the number increases, revealing input scaling as a systematic challenge. We further show that combining self-consistency~\cite{wang2023selfconsistency} with audio permutation yields up to 6.28\% accuracy gains, which increase to 6.74\% when combined with Chain-of-Thought~\cite{cot, cot2} reasoning. These findings shed light on the limitations of LALMs.

Our contributions are threefold: (1) We introduce MUGEN, a comprehensive benchmark for multi-audio understanding in LALMs. (2) We identify key challenges, particularly in non-semantic attributes and input scaling. (3) We study effective training-free strategies to improve multi-audio performance. The MUGEN benchmark is available at \url{https://github.com/danielqwer/MUGEN}.

% The MUGEN benchmark is available at \url{https://huggingface.co/Multi-Audio-Grounding}.

% We will release the benchmark after the review process.

\section{MUGEN Benchmark}
\subsection{Overview}

MUGEN evaluates multi-audio understanding in LALMs through \textbf{35} tasks totaling \textbf{1750} test instances across \textbf{7} dimensions spanning speech, audio, and music. Each task is formulated as a multiple-choice audio grounding problem: given a textual constraint, the model selects the audio candidate that best satisfies it. For example, under the constraint ``select the audio with the angriest emotion," the model must recognize and compare the emotional intensity of each candidate, requiring joint reasoning over multiple audio inputs. Each task provides \textbf{five} audio candidates, and ten tasks additionally include a reference audio (e.g., ``select the option featuring the same speaker as the reference"), resulting in \textbf{six} audio inputs that require reference-conditioned comparison. By enforcing cross-audio comparison and reference-conditioned reasoning, MUGEN probes integrated multi-audio reasoning rather than isolated perception. Table~\ref{tab:statistics} summarizes the dataset statistics.

\begin{table}[t]
  \small
  \centering
  \setlength{\tabcolsep}{6pt}
  \renewcommand{\arraystretch}{0.8}
  \caption{Dataset statistics of MUGEN. Duration and instruction length are reported as mean $\pm$ standard deviation.}
  \vspace{-2mm}
  % \label{tab:statistics}
  \begin{tabular}{l|r}
    \toprule
    Statistic & Value \\
    \midrule
    \# Audio clips & 9,250 \\
    Audio duration (s) & $8.60 \pm 8.79$ \\
    Instruction length (words) & $13.91 \pm 5.46$ \\
    \bottomrule
  \end{tabular}

  \label{tab:statistics}
  \vspace{-15pt}
\end{table}

\begin{table*}[ht]
  \centering
  \caption{Accuracy (\%) and 95\% confidence interval of baselines on MUGEN across seven dimensions: Semantics and Pragmatics (S\&P), Speaker and Demographics (S\&D), Affective and Paralinguistic (A\&P), Temporal Awareness (TA), Acoustic Scene and Event Analysis (AS\&E), Music Analysis (MA), and Compositional Acoustic Reasoning (CA). Overall accuracy shows the micro-average over all tasks. “Low”/“High” denote the thinking levels of Gemini. The best performance within each group is highlighted in bold.}
  \vspace{-5pt}
  \renewcommand{\arraystretch}{0.9}
  \label{tab:model_performance}
  \resizebox{\textwidth}{!}{
    \begin{tabular}{l|c|cccccc|c}
      \toprule
      \textbf{Model}
      & \textbf{S\&P} ($\uparrow$)
      & \textbf{S\&D} ($\uparrow$)
      & \textbf{A\&P} ($\uparrow$)
      & \textbf{TA} ($\uparrow$)
      & \textbf{AS\&E} ($\uparrow$)
      & \textbf{MA} ($\uparrow$)
      & \textbf{CA} ($\uparrow$)
      & \textbf{Overall} ($\uparrow$) \\
      \midrule
      \multicolumn{9}{c}{\textbf{Open-source LALMs}} \\
      \midrule
      \textbf{DeSTA2.5-Audio}
      & 46.67 {\scriptsize $\pm$ 7.98} & 22.00 {\scriptsize $\pm$ 5.74} & 21.75 {\scriptsize $\pm$ 4.04} & 17.20 {\scriptsize $\pm$ 4.68} & \textbf{30.40} {\scriptsize $\pm$ 5.70} & 18.00 {\scriptsize $\pm$ 4.76} & \textbf{28.40} {\scriptsize $\pm$ 5.59} & 24.91 {\scriptsize $\pm$ 2.03} \\
      \textbf{Qwen2.5-Omni}
      & \textbf{70.00} {\scriptsize $\pm$ 7.33} & 12.00 {\scriptsize $\pm$ 4.50} & \textbf{32.50} {\scriptsize $\pm$ 4.59} & 15.20 {\scriptsize $\pm$ 4.45} & 10.80 {\scriptsize $\pm$ 3.85} & \textbf{53.20} {\scriptsize $\pm$ 6.19} & 18.00 {\scriptsize $\pm$ 4.76} & \textbf{28.69} {\scriptsize $\pm$ 2.12} \\
      \textbf{Audio Flamingo 3}
      & 25.33 {\scriptsize $\pm$ 6.96} & 22.50 {\scriptsize $\pm$ 5.79} & 15.00 {\scriptsize $\pm$ 3.50} & 21.20 {\scriptsize $\pm$ 5.07} & 23.20 {\scriptsize $\pm$ 5.23} & 1.20 {\scriptsize $\pm$ 1.35} & 19.20 {\scriptsize $\pm$ 4.88} & 17.43 {\scriptsize $\pm$ 1.78} \\
      \textbf{Voxtral-Mini-3B}
      & 59.33 {\scriptsize $\pm$ 7.86} & 25.50 {\scriptsize $\pm$ 6.04} & 30.00 {\scriptsize $\pm$ 4.49} & 20.00 {\scriptsize $\pm$ 4.96} & 22.40 {\scriptsize $\pm$ 5.17} & 24.80 {\scriptsize $\pm$ 5.35} & 23.60 {\scriptsize $\pm$ 5.26} & 27.83 {\scriptsize $\pm$ 2.10} \\
      \textbf{Voxtral-Small-24B}
      & 64.67 {\scriptsize $\pm$ 7.65} & \textbf{27.50} {\scriptsize $\pm$ 6.19} & \textbf{32.50} {\scriptsize $\pm$ 4.59} & 22.80 {\scriptsize $\pm$ 5.20} & 22.40 {\scriptsize $\pm$ 5.17} & 25.60 {\scriptsize $\pm$ 5.41} & 16.80 {\scriptsize $\pm$ 4.63} & 28.63 {\scriptsize $\pm$ 2.12} \\
      \textbf{Phi-4-multimodal-instruct}
      & 52.00 {\scriptsize $\pm$ 8.00} & 21.00 {\scriptsize $\pm$ 5.65} & 21.00 {\scriptsize $\pm$ 3.99} & \textbf{28.00} {\scriptsize $\pm$ 5.57} & 24.80 {\scriptsize $\pm$ 5.35} & 24.80 {\scriptsize $\pm$ 5.35} & 24.80 {\scriptsize $\pm$ 5.35} & 26.29 {\scriptsize $\pm$ 2.06} \\

      \midrule
      \multicolumn{9}{c}{\textbf{Proprietary LALM}} \\
      \midrule

      \textbf{Gemini-3-pro (Low)}
      & 89.33 {\scriptsize $\pm$ 4.94} & 68.00 {\scriptsize $\pm$ 6.47} & 77.00 {\scriptsize $\pm$ 4.12} & 48.00 {\scriptsize $\pm$ 6.19} & 55.60 {\scriptsize $\pm$ 6.16} & \textbf{69.60} {\scriptsize $\pm$ 5.70} & 69.20 {\scriptsize $\pm$ 5.72} & 67.66 {\scriptsize $\pm$ 2.19} \\
      \textbf{Gemini-3-pro (High)}
      & \textbf{90.67} {\scriptsize $\pm$ 4.65} & \textbf{71.50} {\scriptsize $\pm$ 6.26} & \textbf{77.50} {\scriptsize $\pm$ 4.09} & \textbf{53.60} {\scriptsize $\pm$ 6.18} & \textbf{56.80} {\scriptsize $\pm$ 6.14} & 68.40 {\scriptsize $\pm$ 5.76} & \textbf{72.80} {\scriptsize $\pm$ 5.52} & \textbf{69.60} {\scriptsize $\pm$ 2.16} \\

      \midrule
      \multicolumn{9}{c}{\textbf{Cascaded Systems}} \\
      \midrule
      \textbf{ASR + LLM (Low)}
      & 75.33 {\scriptsize $\pm$ 6.90} & \textbf{20.00} {\scriptsize $\pm$ 5.54} & 27.00 {\scriptsize $\pm$ 4.35} & \textbf{28.80} {\scriptsize $\pm$ 5.61} & \textbf{26.40} {\scriptsize $\pm$ 5.46} & \textbf{26.00} {\scriptsize $\pm$ 5.44} & 18.00 {\scriptsize $\pm$ 4.76} & 29.09 {\scriptsize $\pm$ 2.13} \\
      \textbf{ASR + LLM (High)}
      & \textbf{82.67} {\scriptsize $\pm$ 6.06} & 16.50 {\scriptsize $\pm$ 5.14} & \textbf{27.75} {\scriptsize $\pm$ 4.39} & \textbf{28.80} {\scriptsize $\pm$ 5.61} & 26.00 {\scriptsize $\pm$ 5.44} & \textbf{26.00} {\scriptsize $\pm$ 5.44} & \textbf{22.40} {\scriptsize $\pm$ 5.17} & \textbf{30.06} {\scriptsize $\pm$ 2.15} \\
      \bottomrule
    \end{tabular}
  }
\end{table*}
% \begin{figure}[ht!]
%   \centering
%   % First subfigure
%   \begin{subfigure}{\linewidth}
%     \centering
%     \includegraphics[width=\linewidth]{figures/scaling_results.png}
%     \vspace{-10pt}
%     \caption{Accuracy as the number of audio candidates increases from 2 to 5.}
%     \label{fig:scaling_absolute_value}
%   \end{subfigure}

%   \vspace{0.5em}

%   % Second subfigure
%   \begin{subfigure}{\linewidth}
%     \centering
%     \includegraphics[width=\linewidth]{figures/scaling_retention.png}
%     \vspace{-10pt}
%     \caption{Accuracy at each $n$ normalized by the accuracy at $n=2$.}
%     \label{fig:scaling_retention}
%   \end{subfigure}

%   \caption{Performance under varying numbers of audio candidates (left: without reference; right: with reference).
%     Top: accuracy for $n=2$–$5$ candidate audios.
%   Bottom: accuracy relative to the $n=2$ setting, $\mathrm{Acc}(n)/\mathrm{Acc}(2)$.}

%   \label{fig:scaling}
% \end{figure}

\begin{figure*}[t]
  \centering

  % -------- Left subfigure --------
  \begin{subfigure}{0.47\linewidth}
    \centering
    \includegraphics[width=\linewidth]{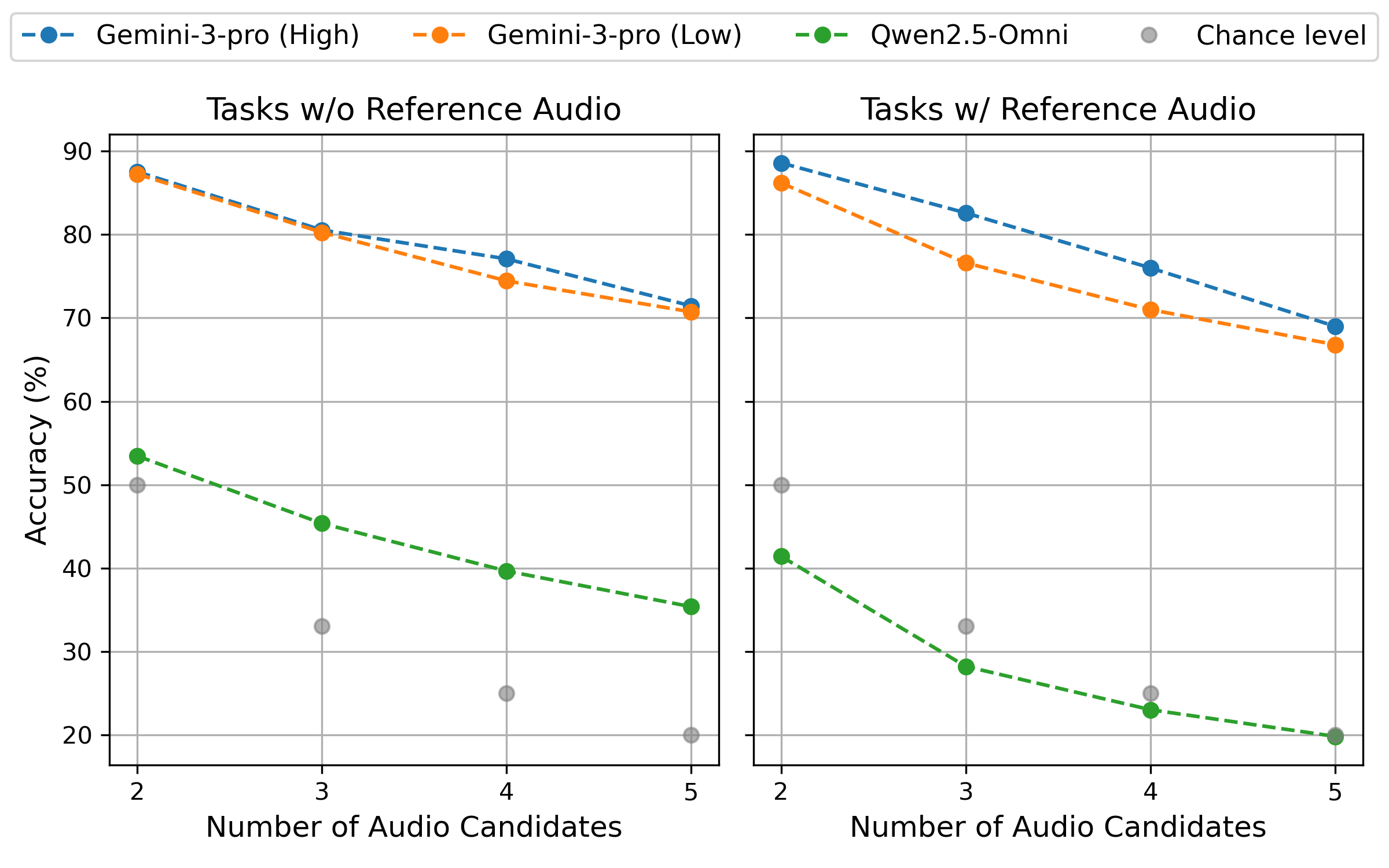}
    \vspace{-4mm}
    \caption{Overall accuracy for $n=2$–$5$ candidate audios.}
    \label{fig:scaling_absolute_value}
  \end{subfigure}
  \hfill
  % -------- Right subfigure --------
  \begin{subfigure}{0.47\linewidth}
    \centering
    \includegraphics[width=\linewidth]{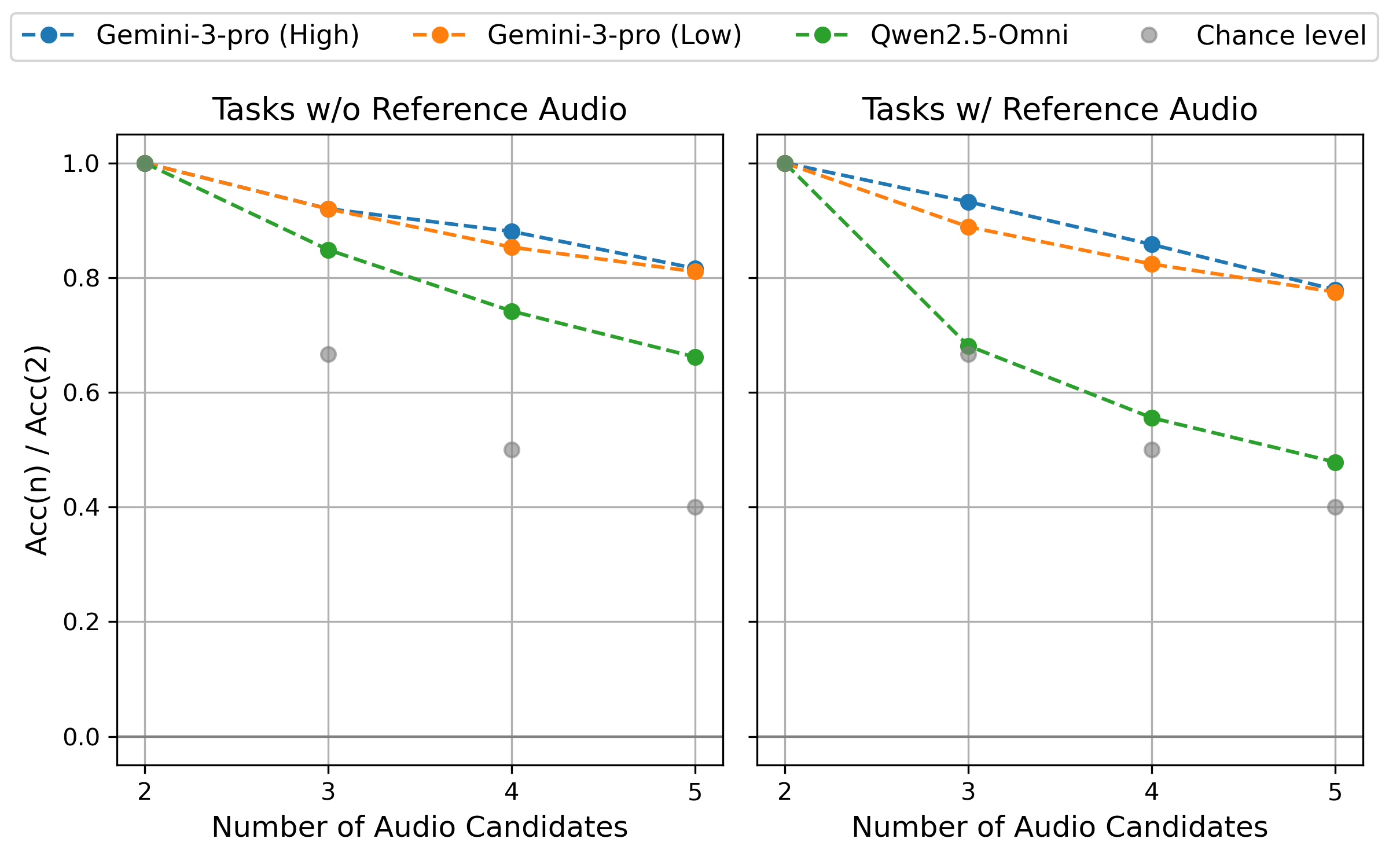}
    \vspace{-4mm}
    \caption{Overall accuracy relative to $n=2$, $\mathrm{Acc}(n)/\mathrm{Acc}(2)$.}
    \label{fig:scaling_retention}
  \end{subfigure}
    \vspace{-2mm}
  \caption{Performance scaling under varying numbers of audio candidates for tasks without (left) and with reference audio (right).}

  \label{fig:scaling}
\end{figure*}

\subsection{Evaluation Dimensions}

MUGEN covers seven complementary dimensions of multi-audio understanding.
\emph{Semantics \& Pragmatics} evaluates comprehension of speech content and contextual meaning, where tasks can in principle be solved by transcribing the audio and reasoning over text.
\emph{Speaker \& Demographics} assesses recognition of speaker-related attributes such as identity cues and accent.
\emph{Affective \& Paralinguistic State} examines non-semantic vocal signals, including emotion and prosody.
\emph{Temporal Awareness} probes sensitivity to temporal properties such as duration and pacing.
\emph{Acoustic Scene \& Event Analysis} measures identification and reasoning over environmental sounds.
\emph{Music Analysis} evaluates recognition of musical attributes such as genre and instrumentation.
\emph{Compositional Acoustic Reasoning} tests integrated reasoning over multiple attributes across dimensions.

\subsection{Data Sources}
We source the majority of auditory data from publicly available corpora, using synthetic generation only when precise attribute control is required. Standard classification tasks covering emotion, speaker identity, language, and music are derived from widely adopted speech and audio benchmarks \cite{6849440, 7178964, ardila-etal-2020-common, bogdanov2019mtg}. For attributes involving specific vocal conditions, we curate samples from specialized academic corpora~\cite{rudzicz2012torgo, Wilkins2018VocalSetAS, bastianelli-etal-2020-slurp, yamagishi_cstr_2019, cummins2006chains, zhao2018l2arctic, 10.1007/11677482_3, 10902687, thiemann_demand_2013, piczak2015dataset, nguyen23_interspeech} as well as high-quality scripted speech collections\footnote{\url{https://datacollective.mozillafoundation.org/datasets/cmkfm9fbl00nto0070sdcrak2}}. Finally, prosodic and semantic tasks that require strict ground-truth alignment are constructed using open-source speech synthesis models \cite{wang2025maskgct, hu2026qwen3} or GPT-4o mini TTS. All synthesized samples are manually verified to ensure data quality and attribute fidelity.

Regarding task construction, all textual instructions are manually authored to provide clear and unambiguous selection constraints. To formulate the multiple-choice options, we first establish the ground-truth audio that satisfies the target objective. The remaining distractor candidates are then systematically selected or synthesized to exhibit contrasting variations of the specific auditory attribute under evaluation. This rigorous candidate formulation ensures that the model must perform precise cross-audio comparison to distinguish the correct option, preventing reliance on superficial shortcuts.

\section{Experimental Setups}
\subsection{Baselines}
\label{sec:baselines}
We evaluate a diverse set of advanced LALMs to assess current progress in multi-audio understanding, including open-source models like DeSTA2.5-Audio~\cite{desta25}, Qwen2.5-Omni-7B~\cite{qwen25omni}, Audio Flamingo 3~\cite{af3}, Voxtral-Mini-3B, Voxtral-Small-24B~\cite{voxtral}, and Phi-4-Multimodal-Instruct~\cite{phi4}, as well as the proprietary Gemini-3-pro~\cite{gemini3pro} under different thinking levels. We run the open-source models with the vLLM framework~\cite{vllm} and include a cascade baseline (ASR+LLM) that transcribes audio with Whisper-large-v3~\cite{whisper} and answers with Gemini-3-pro, to estimate how much of our tasks can be solved with semantic content alone. We use greedy decoding for all models except Voxtral and Gemini-3-pro. For Voxtral, we adopt the decoding configuration strongly recommended by the original authors (temperature 0.2 and top-$p$ 0.95). For Gemini-3-pro, we use temperature 1, as highly recommended by the official team.

% We use greedy decoding for all models except Voxtral and Gemini-3-pro. For the Voxtral models, we follow the decoding configuration recommended by the original authors (temperature $0.2$ and top-$p$ $0.95$). For Gemini-3-pro, we adopt the recommended setting with temperature $1$.

\subsection{Evaluation Metric}
All tasks are formulated as multiple-choice question answering, with accuracy as the metric.
To extract answers from free-form outputs, we adopt LLM-as-a-judge~\cite{llm-judge}, using Claude Haiku 4.5\footnote{claude-haiku-4-5-20251001} with temperature 0 for reproducibility.

The judge evaluates each prediction against the ground truth under predefined rubrics: each sample has exactly one correct option, and selecting none or multiple options is marked incorrect. It also provides a brief explanation and outputs the decision in a fixed format for post-processing, with human-annotated examples for calibration.
On 400 randomly sampled instances, we observe 99\% agreement with human annotators, supporting the reliability of automatic evaluation.

\section{Evaluation Results}

\subsection{Main Results}
\label{sec:main_results}

Table~\ref{tab:model_performance} presents the performance of baseline models on the MUGEN benchmark. We discuss the main observations below.

\textbf{Current LALMs remain limited in multi-audio understanding.}
We compare cascaded ASR+LLM systems with end-to-end LALMs. While ASR+LLM performs competitively on semantic tasks, its performance degrades noticeably on non-semantic dimensions, suggesting that semantic-based reasoning alone is insufficient for many tasks in our benchmark. However, open-source end-to-end LALMs, despite direct access to acoustic signals, achieve overall performance only comparable to cascaded systems. This indicates that open-source models remain limited in multi-audio understanding and fail to extend their strong abilities in single-audio scenarios~\cite{sakura, airbench, audiobench} to multi-audio settings. In contrast, the proprietary model consistently and substantially outperforms all open-source models across dimensions. Nevertheless, even the strongest model remains far from perfect, indicating that multi-audio understanding remains an open challenge even for proprietary systems.

\textbf{LALMs exhibit a clear imbalance between semantic and non-semantic dimensions.}
All LALMs perform consistently better on the semantic dimension than on others, indicating stronger semantic reasoning while non-semantic perceptual reasoning remains underdeveloped. These results reveal an uneven capability distribution and highlight the need for more balanced modeling of diverse auditory attributes in multi-audio settings.

\textbf{Systematic blind spots persist across non-semantic dimensions.}
Among the non-semantic dimensions, some are consistently more challenging for most models. In particular, reasoning over and comparing temporal information remains difficult, with substantially lower performance even for proprietary models. We also observe model-specific weaknesses, such as music analysis for Audio Flamingo 3 and compositional acoustic reasoning for Voxtral-Small-24B. These results reveal concrete gaps in current LALMs and suggest that targeted training strategies or architectural adjustments may be required to strengthen multi-audio understanding.

\subsection{Performance Scaling with the Number of Audio Inputs}
While Sec.~\ref{sec:main_results} highlights the overall limitations of LALMs in multi-audio understanding, it remains unclear how performance scales with the number of audio inputs. Thus, we construct reduced variants of MUGEN by progressively decreasing the number of audio candidates from five to two. This procedure controls input size while preserving the task formulation.

For each instance, we remove one non-ground-truth option while keeping the instruction unchanged. Ranking-based tasks are excluded, as reducing the option set may invalidate instructions (e.g., “select the third longest audio” becomes ill-defined when fewer than three audios are provided). We therefore conduct this analysis on the remaining 32 tasks.

We evaluate Qwen2.5-Omni and Gemini-3-pro (both thinking levels), the strongest models in Table~\ref{tab:model_performance}. As shown in Figures~\ref{fig:scaling_absolute_value} and~\ref{fig:scaling_retention}, accuracy declines as the number of candidate audios increases from two to five. To compare degradation across models, we report accuracy at each $n$ relative to the two-candidate setting, $\mathrm{Acc}(n)/\mathrm{Acc}(2)$ (Figure~\ref{fig:scaling_retention}), where $n$ denotes the number of candidate audios (excluding the reference when present). Qwen2.5-Omni shows a sharper performance decline than Gemini-3-pro. With five candidates, it preserves only about 66\% and 48\% of its two-candidate accuracy in tasks without and with reference audio, respectively, whereas Gemini-3-pro retains around 80\% in both cases. The two thinking levels of Gemini-3-pro follow similar trends, indicating that increased thinking depth does not mitigate scaling degradation. Notably, the decline already emerges with only three additional audio inputs. \textbf{These findings identify input scaling as a systematic challenge for current LALMs.}

This analysis characterizes LALM behavior under controlled input expansion. Such evaluation is difficult with existing benchmarks, where the number of audio inputs is limited. By enabling systematic variation of input size, MUGEN facilitates fine-grained analysis under increasing input complexity.

\begin{table}[t]
\centering 
\caption{Performance of improvement strategies on MUGEN. Main values report accuracy (\%), while small numbers indicate absolute gains over the original performance (\textit{Orig.}).}
\vspace{-5pt}
\label{tab:improve_table_diff}
\setlength{\tabcolsep}{2pt}
\resizebox{\columnwidth}{!}{
\begin{tabular}{l ccc}
\toprule
\textbf{Method} & \textbf{Qwen2.5-Omni} & \textbf{Gemini-3-pro (Low)} & \textbf{Gemini-3-pro (High)} \\
\midrule
\textit{Orig.} & 28.69 & 67.66 & 69.60 \\
\textit{CoT} & 28.57\rlap{\,{\scriptsize\textcolor{red}{-0.12}}} & 67.89\rlap{\,{\scriptsize\textcolor{green!60!black}{+0.23}}} & 70.46\rlap{\,{\scriptsize\textcolor{green!60!black}{+0.86}}} \\
\midrule
\textit{SC} & 28.00\rlap{\,{\scriptsize\textcolor{red}{-0.69}}} & 70.06\rlap{\,{\scriptsize\textcolor{green!60!black}{+2.40}}} & 72.74\rlap{\,{\scriptsize\textcolor{green!60!black}{+3.14}}} \\
\textit{APSC} & 30.69\rlap{\,{\scriptsize\textcolor{green!60!black}{+2.00}}} & 73.94\rlap{\,{\scriptsize\textcolor{green!60!black}{+6.28}}} & 74.97\rlap{\,{\scriptsize\textcolor{green!60!black}{+5.37}}} \\
\midrule
\textit{SC+CoT} & 28.34\rlap{\,{\scriptsize\textcolor{red}{-0.35}}} & 69.94\rlap{\,{\scriptsize\textcolor{green!60!black}{+2.28}}} & 73.49\rlap{\,{\scriptsize\textcolor{green!60!black}{+3.89}}} \\
\textit{APSC+CoT} & \textbf{31.31}\rlap{\,{\scriptsize\textcolor{green!60!black}{+2.62}}} & \textbf{74.40}\rlap{\,{\scriptsize\textcolor{green!60!black}{+6.74}}} & \textbf{75.26}\rlap{\,{\scriptsize\textcolor{green!60!black}{+5.66}}} \\
\bottomrule
\end{tabular}
}

\vspace{-10pt}
\end{table}

\section{Improvement Strategies}

\subsection{Methodology}
We investigate several training-free strategies to improve the multi-audio understanding of LALMs.
To encourage structured deliberation before answer prediction, we examine \emph{Chain-of-Thought} (CoT)~\cite{cot, cot2}, which elicits intermediate reasoning steps (e.g., ``Let's think step by step'').
Alongside CoT, we evaluate \emph{Self-Consistency} (SC)~\cite{wang2023selfconsistency} to improve robustness.
Instead of relying on a single decoding output, SC generates multiple sampled responses and aggregates them via majority voting.

Building upon SC, we investigate \emph{Audio-Permutational Self-Consistency} (APSC) for multi-audio scenarios.
Before each inference, we randomly permute the order of audio candidates to reduce positional sensitivity and over-reliance on a specific audio position~\cite{lin2025hearing}.
The model generates responses under these permutations, which are mapped back to the original indexing system and aggregated via majority voting.

For implementation, SC and APSC use a temperature of 0.2 for Qwen2.5-Omni, while Gemini-3-pro follows the recommended configuration in Sec.~\ref{sec:baselines}. Both methods generate 10 responses for majority voting. In SC, all responses are sampled under the original audio order, whereas APSC generates one response for each of the 10 permutations, maintaining the same number of generations and comparable computational cost.

% For implementation, SC and APSC utilize a temperature of 0.2 for Qwen2.5-Omni and 1.0 for Gemini-3-pro, generating 10 responses for majority voting.
% In SC, all responses are sampled under the original audio order.
% Conversely, APSC generates one response for each of the 10 permutations, ensuring an identical number of generations and comparable computational cost.

\subsection{Results}

\textbf{Chain-of-Thought reasoning fails to resolve underlying auditory perceptual bottlenecks.}
Table \ref{tab:improve_table_diff} details the performance of various improvement strategies.
All models consistently benefit most from APSC, followed by SC, while CoT yields minimal or even negative impacts.
Specifically, CoT leads to only marginal improvements for proprietary models and slightly degrades Qwen2.5-Omni’s accuracy.
This suggests that the primary challenge lies in acoustic perception and cross-audio comparison, rather than a deficiency in logical reasoning.
Simply prompting models to deliberate over text cannot compensate for its inability to accurately differentiate auditory features.

\textbf{Audio permutation further enhances self-consistency by diversifying the presentation order of audio candidates.}
Building upon the improvements of standard SC, APSC consistently achieves superior performance.
By processing the audio options under varied presentation orders, APSC helps models form a more robust aggregated prediction, indicating that sensitivity to input arrangement can be effectively mitigated. Furthermore, combining this permutation strategy with CoT yields peak performance across all models, including an absolute accuracy gain of up to 6.74\% for Gemini-3-pro (Low). However, while this method achieves significant improvements, producing multiple generations per instance incurs considerable computational overhead. Nevertheless, these findings suggest that audio permutation serves as a beneficial strategy for improving multi-audio comprehension in LALMs.

\section{Conclusion}

We introduce MUGEN, a comprehensive benchmark for evaluating multi-audio understanding in LALMs. Our systematic evaluation reveals critical blind spots in state-of-the-art models: while proficient in semantic tasks, their performance degrades severely on non-semantic attributes and scales poorly with concurrent audio inputs. Furthermore, we demonstrate that common CoT is insufficient for these multi-audio challenges. Instead, mitigating inherent positional biases through APSC yields substantial accuracy improvements. By exposing these foundational limitations and validating effective inference strategies, MUGEN establishes a vital stepping stone for advancing complex auditory comprehension in future LALMs.

\section{Acknowledgement}
% We acknowledge the computational and storage support provided by the National Center for High-performance Computing (NCHC) of the National Applied Research Laboratories (NARLabs) in Taiwan.

This work was supported by the Ministry of Education (MOE), Taiwan, through the NTU Artificial Intelligence Center of Research Excellence (NTU AI-CoRE) under the Taiwan Centers of Excellence in Artificial Intelligence project. We also thank the National Center for High-performance Computing (NCHC) of the National Applied Research Laboratories (NARLabs), Taiwan, for providing computational and storage resources.

\section{Generative AI Use Disclosure}

In this paper, generative AI tools were used for writing refinement and language polishing. Additionally, large language models were employed as judges in our automatic evaluation.

\bibliographystyle{IEEEtran}
\bibliography{mybib}

\end{document}